\documentclass[aps,showpacs,twocolumn,superscriptaddress]{revtex4}
\usepackage{graphicx}
\usepackage{dcolumn}

\begin{document}

\title{Radial sensitivity of kaonic atoms and strongly bound $\bar K$ states}

\author{N.~Barnea}
\email{nir@phys.huji.ac.il}
\affiliation{Racah Institute of Physics, The Hebrew University,
Jerusalem 91904, Israel}

\author{E.~Friedman}
\email{elifried@vms.huji.ac.il}
\affiliation{Racah Institute of Physics, The Hebrew University,
Jerusalem 91904, Israel}

\date{\today}

\begin{abstract}
The strength of the low energy $K^-$-nucleus real potential 
 has recently received renewed attention in view of
experimental evidence for the 
possible existence of strongly bound  $K^-$ states. Previous
fits to kaonic atom data led to either `shallow' or  to `deep' potentials, 
where
only the former are in agreement with chiral approaches but
only the latter can produce strongly bound states. Here we explore the
uncertainties of the $K^-$-nucleus optical potentials,
obtained from fits to kaonic atom data, using 
the functional derivatives of the best-fit $\chi ^2$ values with respect to the
potential. We find that only the deep 
type of potential provides information which is applicable
to the $K^-$ interaction in the nuclear interior.

\end{abstract}

\maketitle

\section{Motivation and Background}
\label{sec:mot}
Since the early days of kaonic atom experiments it was known \cite{Kre71,KSW71}
that due to the strength of the $K^-$ nuclear absorption 
the real part of the $K^-$-nucleus potential 
played a secondary role compared to the imaginary part of the potential
and consequently
it could not be determined
uniquely from fits to the data for a single target nucleus.
 More recent `global' fits
to large sets of data encompassing the whole of the periodic table
showed~\cite{BFG97} that although traditional `$t\rho $' potentials 
yield reasonably good fits to the data, the use of phenomenological
density-dependent
$t(\rho )$ amplitudes leads to significantly better fits. When
extrapolated into the interior of nuclei 
the real part of the potentials is typically
180 MeV deep for the density-dependent variety
whereas for the `$t\rho $' potentials it is  typically less than 
100~MeV. We note that chiral-motivated $K^-$-nucleus potentials 
\cite{ROs00,CFG01} are shallower than the phenomenological
`$t\rho $' potentials.

\begin{figure}[t]
\centerline{\includegraphics[height=7cm]{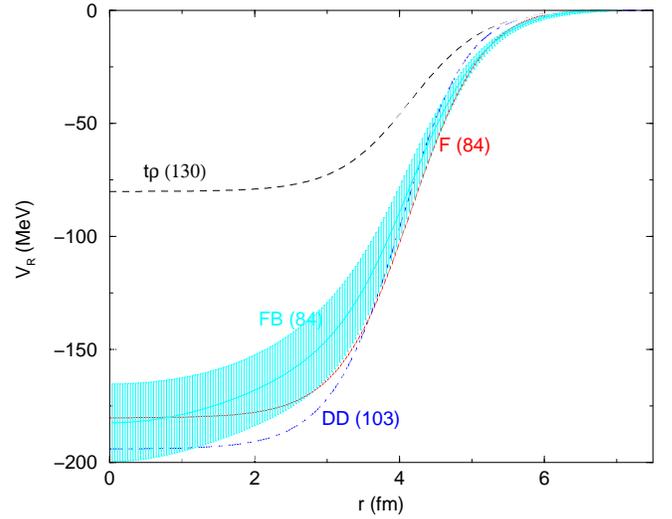}}
\caption{Real part of the $K^-$-Ni optical potential for various models
and the values of $\chi ^2$ for 65 data points in parentheses, see text.}
\label{fig:NiVRFB}
\end{figure}

Figure \ref{fig:NiVRFB} shows, as an example, 
the real part of the $K^-$ optical potential
for Ni obtained from global fits to kaonic atoms data using several
phenomenological models for the interaction. The simplest `$t\rho $'
approach where the real and imaginary parts of the effective $t$-matrix 
are determined from fits to the data yields a $\chi ^2$ of 130 for the
65 data points used in the fit. Adding an adjustable non-linear term
leads to the deep potential `DD' of Ref.~\cite{BFG97} with a $\chi ^2$
of 103 and a greatly increased depth. Also shown in the figure is 
another potential using a geometrical approach to the density-dependence
of $t$ (the `F' potential of Ref.\cite{MFG06}), leading to $\chi ^2$ of 84.
The similarity of the two deep potentials and the great difference compared
to the shallow one are clearly observed. The other curve 
`FB' with its error
band, also of $\chi^2 = 84$,
is discussed below.
A consequence of the depth of the potential is
the ability to support a strongly bound state, 
a question that was
highlighted again recently by 
experimental reports on candidates for $\bar K$-nuclear 
bound states in the range of binding energy
$B_{\bar K} \sim 100 - 200$~MeV~\cite{SBF04,SBF05,KHA05,ABB05}.
However, very recently a possible explanation of  the experimental 
results of FINUDA \cite{ABB05} was given in terms of $\Lambda p$
final state interaction \cite{MOR06} and the observed width 
was shown to be at odds
with recent Faddeev calculations \cite{SGM06}.

 Obviously, the `$t\rho $'-type of
potential cannot generate strongly bound nuclear states
in the energy range $B_{\bar K} \sim 100 - 200$ MeV, whereas the deeper
potentials (`DD' and `F') might do.  
In the present work we address the question of how well is the real
part of the 
$K^-$-nucleus potential determined, with its ability to support
strongly bound states as the topic of interest.

Estimating the uncertainties of hadron-nucleus potentials as function
of position is not a simple task. For example, in the `$t\rho $'
approach the shape of the potential is determined by the nuclear
density distribution and the uncertainty in the strength parameter,
as obtained from $\chi ^2$ fits to the data,
implies a fixed {\it relative} uncertainty at all radii, which is, of course,
baseless. Details vary when more elaborate
forms such as `DD' or `F' are used, but one is left essentially
with analytical continuation into the nuclear interior of potentials
that might be well-determined only close to the nuclear surface. 
`Model-independent' methods have been used in analyses of elastic scattering
data for various particles \cite{BFG89} 
to alleviate this problem. However, applying e.g. the 
Fourier-Bessel (FB) method in global analyses of kaonic atom data end up
in too few terms in the series, thus making the uncertainties unrealistic
in their dependence on position. This is illustrated in Fig. \ref{fig:NiVRFB}
by the `FB' curve, obtained by adding a Fourier-Bessel series to a `$t\rho $'
potential. Only three terms in the series are needed to 
achieve a $\chi ^2$ of 84
and the potential becomes deep, in agreement with the other two `deep'
solutions. The error band obtained from the FB method \cite{BFG89}
is, nevertheless,  unrealistic because only three FB 
terms are used. However, an increase
in the number of terms is found to be unjustified numerically in this case.

In the present work we adopt a functional-derivative approach in order
to get more realistic position-dependence  
of the uncertainties of $K^-$-nucleus potentials.
 The method is applied to the two types of
potentials (`shallow' and `deep') mentioned above and it is shown that
deep potentials within nuclei are reliably obtained from fits to
experimental results for kaonic atoms.

\section{Method}
\label{sec:met}

The radial sensitivity of exotic atom data was addressed before 
\cite{BFG97} with the help of a `notch test', introducing a local
perturbation into the potential and studying the changes in the
fit to the data as function of position of the perturbation. The
results gave at least a semi-quantitative information on what are
the radial regions which are being probed by the various types of
exotic atoms. In fact, the difference in that respect
between deep and shallow kaonic 
atom potentials could be observed. However, the extent of the perturbation
was somewhat arbitrary and in the present work we report on
extending that approach to a mathematically well-defined limit.

In order to study the radial sensitivity of {\it global} 
fits to kaonic atom data,
it is instructive to define the radial position parameter in a `natural' way 
using, e.g.,  the
known charge distribution for each nuclear species in the data base. 
The radial position $r$
is then defined as $r=R_c+\eta a_c$, where
$R_c$ and $a_c$ are the radius and diffuseness parameters, respectively, of a 
two-parameter Fermi (2pF) charge distribution \cite{FBH95}. 
In that way $\eta$ becomes the relevant radial parameter 
when handling together data for several nuclear species along the 
periodic table.  
The value of $\chi ^2$ can be regarded now as a functional 
of a global optical potential $V(\eta)$, i.e. $\chi ^2=\chi^2[V(\eta)]$.
The parameter $\eta$ is a {\it continuous} variable, however it is instructive
to start the discussion of variations by assuming that the global optical
potential is defined on a discrete set of grid points $\{\eta_i, i=1\ldots
N\}$, so that $\chi ^2$ depends on a set of $N$ parameters $V_i=V(\eta_i)$. The
variation of $\chi ^2$ due to a small change in these parameters is simply
\begin{equation}\label{dchi2_etai}
 d\chi^2 = \sum_{i=1}^{N} \frac{\partial \chi^2}{\partial V_i} d V_i \;.
\end{equation}
The equivalent expression for the {\it continuous} function $V(\eta)$ can be
obtained 
by taking the limit for a very dense grid, leading to \cite{FD_wikipedia} 
\begin{equation}\label{dchi2}
 d\chi^2 = \int d\eta \frac{\delta \chi^2}{\delta V(\eta)} \delta V(\eta) \;,
\end{equation}
where 
\begin{equation}
\frac{\delta \chi^2[V(\eta)]}{\delta V(\eta')}
= \lim_{\sigma \rightarrow 0}\lim_{\epsilon \rightarrow 0 }
\frac{\chi^2[V(\eta)+\epsilon\delta_{\sigma}(\eta-\eta')]-\chi^2[V(\eta)]}
     {\epsilon}\;
\end{equation}
is the functional derivatives (FD) of $\chi^2[V]$.
The notation $\delta_{\sigma}(\eta-\eta')$ stands for an approximated
$\delta$-function. 
From Eq.~(\ref{dchi2}) it is seen that the FD determines the effect of a local
change in the optical potential on $\chi^2$. Conversely it can be said that
the optical potential sensitivity to the experimental data is determined by
the magnitude of the FD. 
In practice the calculation of the FD was carried out by multiplying the
best fit potential by a factor
\begin{equation}
 f=1+\epsilon \delta_{\sigma}(\eta-\eta')
\end{equation}
using a normalized Gaussian with a range parameter $\sigma$ for the 
smeared $\delta$-function,
\begin{equation}
\delta_{\sigma}(\eta-\eta')=\frac{1}{\sqrt{2\pi}\sigma}e^{-(\eta-\eta')^2/2\sigma^2}.
\end{equation}
For finite values of $\epsilon$ and $\sigma$ the FD can be approximated by
\begin{equation}
\frac{\delta \chi^2[V(\eta)]}{\delta V(\eta')}
\approx \frac{1}{V(\eta')}
\frac{\chi^2[V(\eta)(1+\epsilon\delta_{\sigma}(\eta-\eta'))]
      -\chi^2[V(\eta)]}
     {\epsilon}\;.
\end{equation}
The parameter $\epsilon $ was used for a {\it fractional} 
change in the potential
and the limit $\epsilon \to 0$ was obtained numerically for several
values of $\sigma $ and then extrapolated to $\sigma =0$. 
Good numerical stability and
good convergence were obtained in all cases.
Here $\eta$ and $\sigma$ are dimensionless variables.

\section{RESULTS AND CONCLUSIONS}

The $K^-$-nucleus potentials used in the present work are taken from 
recent global fits \cite{MFG06} to kaonic atom  data from $^6$Li
to U, a total of 65 data points. A two-parameter fit with a $t\rho $
potential yields a total $\chi ^2$ value of 130 whereas a four-parameter
fit with  $t(\rho )$ amplitudes yields $\chi ^2$=84 for the 65 data points.
In calculating FD for global fits radial positions were defined in terms
of units of diffuseness relative to the charge radius, as described above.
For several of the lightest nuclei harmonic oscillator densities are
more appropriate and had  indeed been used 
in the fits of Ref.~\cite{MFG06}. 
These nuclei have been excluded from the calculations
of FD for the global fits so as to have full consistency in the use of 
2pF density distributions. That left 50 data points in the calculations
of the FD.

\begin{figure}[t]
\centerline{\includegraphics[height=7cm]{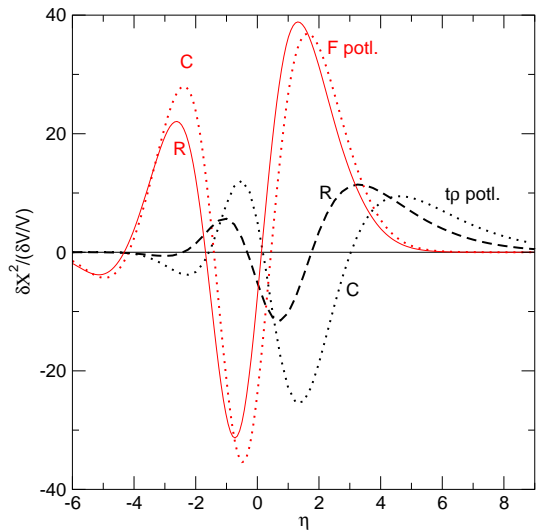}}
\caption{Functional derivatives of $\chi ^2$ with respect to
 the full complex (C) and real (R) potential as function of $\eta$, where
$r=R_c+\eta a_c$, with $R_c$ and $a_c$ the radius and diffuseness
parameters, respectively, of the charge distribution. 
Results are shown for
the $t\rho $ potential and for the $t(\rho )$ `F' potential
of Ref. \cite{MFG06}.}
\label{fig:FDglbR}
\end{figure}

Figure \ref{fig:FDglbR} shows the FD for {\it relative} variations in
the real potential and in the full complex potential. Inspecting the
FD with respect to the imaginary potential (not shown) reveals additivity
of the FD, hence the differences between the FD for the complex potential
and for the real potential in the figure are the corresponding FD with respect
to the imaginary potential. We note that the FD for the deep `F' potential
is dominated by the real part of the potential whereas for the `$t\rho$'
potential the two parts make similar contributions to the FD.
The appearance of regions with negative FD need not be
surprising because it means that some local variations in the shape of the
potentials may cause further reduction in the values of $\chi ^2$. Such
variations are impossible in the $t\rho $ potential. However, such variations
are included when the more flexible `F' potential
is introduced, and the minimisation process essentially follows implicitly
the various FD. We avoid at
present making a quantitative use of the local values of the FD, rather 
we identify with the help of the FD the radial regions to which the 
kaonic atom data are sensitive.

From Fig. \ref{fig:FDglbR} it can be inferred that the sensitive region
for the real $t\rho $ potential is between $\eta =-1.5$ and $\eta =6$
whereas for the F potential it is between $\eta =-3.5$ and $\eta =4$.
Recall that $\eta =-2.2$ correspond to 90\% of the central charge density
and $\eta =2.2$ correspond to 10\% of that density. It therefore
becomes clear that within the $t\rho $ potential there is no sensitivity
 to the interior of the nucleus whereas with the $t(\rho )$ `F' potential,
which yields greatly improved fit to the data, there is sensitivity
to regions within the full nuclear density. 

\begin{figure}[t]
\centerline{\includegraphics[height=8cm]{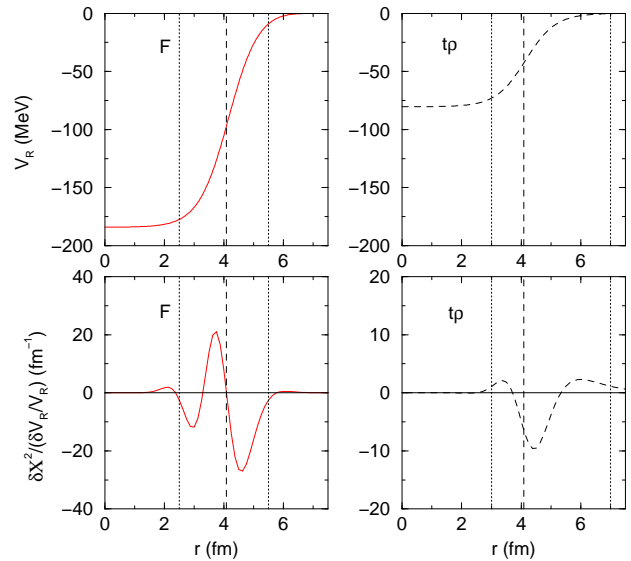}}
\caption{Real potentials (top) and FD (bottom) for the `F' potential
(left) and the $t\rho $ potential (right) for $K^-$ interaction with Ni. 
The regions between the vertical dotted  lines
indicate where the potentials are determined reliably, see text.}
\label{fig:Niall}
\end{figure}

\begin{figure}[t]
\centerline{\includegraphics[height=8cm]{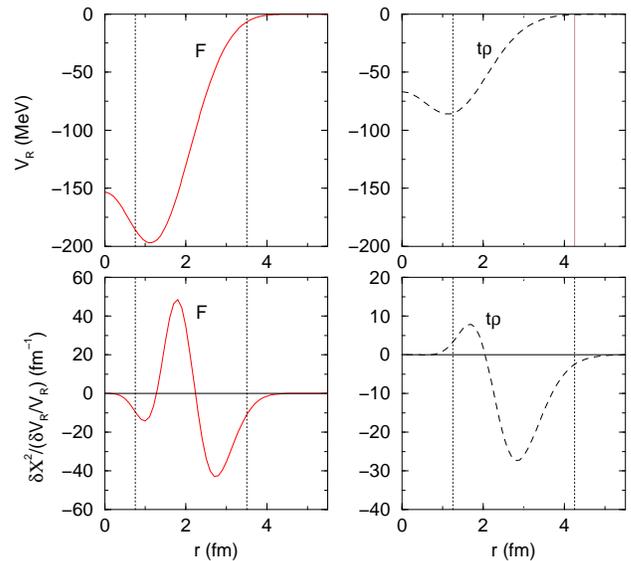}}
\caption{Same as fig. \ref{fig:Niall} but for $^{12}$C.}
\label{fig:Call}
\end{figure}

Figure \ref{fig:Niall} shows similar results for Ni, this time the
radial variable is $r$, the radial position. 
Here $\sigma$ is in units of fm and therefore the FD is in units 
of fm$^{-1}$.
On the left are shown
the real potential (top) and the FD (bottom) for the real `F' potential
and on the right are shown the corresponding quantities for the
$t\rho $ potential. 
The regions between the two vertical dotted lines indicate where 
variation of the potential will affect the fit to the data, thus
suggesting the regions where the potentials are determined  
by experiment. The vertical dashed line indicates 50\% of the central
charge density. It is evident that for the `F' potential 
 the sensitivity extends to depths 
of 180 MeV at
radii where the  density is essentially the full nuclear density.
Very similar results are obtained also for heavier nuclei, where, e.g.
for Pb the distinction between the $t\rho $ and the `F' potential is even
greater than for Ni, with respect to the sensitivity of the experiment to 
depths and densities.

Finally, in Fig. \ref{fig:Call} are shown similar results for $^{12}$C,
which is one of the targets studied experimentally \cite{ABB05}
and theoretically \cite{MFG06}. Again it is seen that with the `F'
potential the sensitive region is at smaller radii and higher densities
compared to that for the $t\rho $ potentials.

In summary, it is found that the `deep' $\bar K$-nucleus potentials, 
which yield excellent
fits to all the kaonic atom data, are determined reliably 
to depths of  150-180 MeV at regions of almost the full
nuclear matter density and consequently they may be reliably
extrapolated to the nuclear interior. In contrast the `shallow'
type of potentials, which yield inferior fits to the data,
are well determined only at the nuclear surface and one cannot 
infer from these what is the depth of the $\bar K$-nucleus potential
in the nuclear interior. The different sensitivities result from
the potentials themselves: the additional attraction provided
by the deep potentials enhances the {\it atomic} wavefunctions within
the nucleus \cite{BFG97} thus creating the sensitivity at smaller radii.
We conclude that optical potentials derived from the observed
strong-interaction effects in kaonic atoms are sufficiently deep to
support strongly-bound antikaon states, but it does not necessarily
imply that such states do exist. Moreover, the discrepancy between
the very shallow chiral-motivated potentials \cite{ROs00,CFG01} and
the deep phenomenological potentials remains an open problem.

   \section*{Acknowledgments}
This work was supported in part
by the Israel Science Foundation grant 757/05.
We thank A. Gal for many discussions.

\end{document}